\documentclass[aps,prl,twocolumn,superscriptaddress,showpacs]{revtex4}
\usepackage{amsmath,amssymb,graphicx,color,textcomp}

\begin{document}

\title{Spin-to-Orbital Angular Momentum Conversion in Semiconductor Microcavities}
\author{F.~Manni}
\email{francesco.manni@epfl.ch}
\author{K.~G.~Lagoudakis}
\author{T.~K.~Para\"iso}
\author{R.~Cerna}
\author{Y.~Leger}
\affiliation{Institute of Condensed Matter Physics, Ecole Polytechnique F\'{e}d%
\'{e}rale de Lausanne (EPFL), CH-1015, Lausanne, Switzerland}
\author{T.~C.~H.~Liew}
\affiliation{Institute of Theoretical Physics, Ecole Polytechnique F\'{e}d\'{e}rale de
Lausanne (EPFL), CH-1015 Lausanne, Switzerland}
\author{I.~A.~Shelykh}
\affiliation{Science Institute, University of Iceland, Dunhagi-3, IS-107, Reykjavik,
Iceland}
\author{A.~V.~Kavokin}
\affiliation{Physics and Astronomy School, University of Southampton, Highfield,
Southampton, SO171BJ, UK}
\author{F.~Morier-Genoud}
\affiliation{Institute of Condensed Matter Physics, Ecole Polytechnique F\'{e}d%
\'{e}rale de Lausanne (EPFL), CH-1015, Lausanne, Switzerland}
\author{B. Deveaud-Pl\'edran}
\affiliation{Institute of Condensed Matter Physics, Ecole Polytechnique F\'{e}d%
\'{e}rale de Lausanne (EPFL), CH-1015, Lausanne, Switzerland}

\begin{abstract}
We experimentally demonstrate a technique for the generation of optical beams carrying orbital angular momentum using a planar semiconductor microcavity. Despite being isotropic systems, the transverse electric - transverse magnetic (TE-TM) polarization splitting featured by semiconductor microcavities allows for the conversion of the circular polarization of an incoming laser beam into the orbital angular momentum of the transmitted light field. The process implies the formation of topological entities, a pair of optical half-vortices, in the intracavity field.
\end{abstract}

\date{\today}
\maketitle

% 71.36.+c: Polaritons
% 42.50.Tx; Optical angular momentum and its quantum aspects
% 42.25.-p: Wave optics
% 42.60.Jf:	Beam characteristics: profile, intensity, and power; spatial pattern formation
% 42.79.-e: Optical elements, devices, and systems

% 78.67.-n: Optical properties of low-dimensional, mesoscopic, and nanoscale materials and structures

It is well-known that photons can carry both an intrinsic spin and orbital angular momentum~\cite{FrankeArnold2008}. Both the intrinsic spin~\cite{Beth1936} and the orbital angular momentum~\cite{He1995} can generate a torque on macroscopic objects, which could provide the optical drive of micromachines~\cite{Grier2003}, and Doppler shifts of spinning bodies~\cite{Garetz1997,Courtial1998}. It has also been shown that orbital angular momentum can be coherently transferred to atoms~\cite{Andersen2006}, allowing in principle the storage of high dimensional quantum information. In fact, whilst the spin angular momentum is restricted to $\pm\hbar$ for each photon, the orbital angular momentum can take any multiple of $\hbar$. Moreover, again in a view of quantum computation applications, entanglement between orbital angular momentum states has been experimentally demonstrated~\cite{Mair2001}. Further applications of beams carrying orbital angular momentum appear in microscopy~\cite{FrankeArnold2008} and ultra-sensitive interferometry suitable for gravitational wave detection~\cite{Granata2010}.

The first-ever implementation of a laser beam carrying orbital angular momentum was achieved by shining a standard laser beam through a system of lenses~\cite{Allen1992}. Later developments showed that individual optical components
such as spiral waveplates~\cite{Beijersbergen1994,Oemrawsingh2004} or holograms with fork like dislocations~\cite{Bazhenov1990} could achieve the same task. More recently, more compact components based on birefringent materials have been
used~\cite{Marrucci2006,Brasselet2009}. It is notable that all known techniques for the creation of beams with orbital angular momentum require an optically inhomogeneous and/or anisotropic material or strong focusing~\cite{Zhao2007}.
\begin{figure}[tb]
\includegraphics[width=0.5\textwidth]{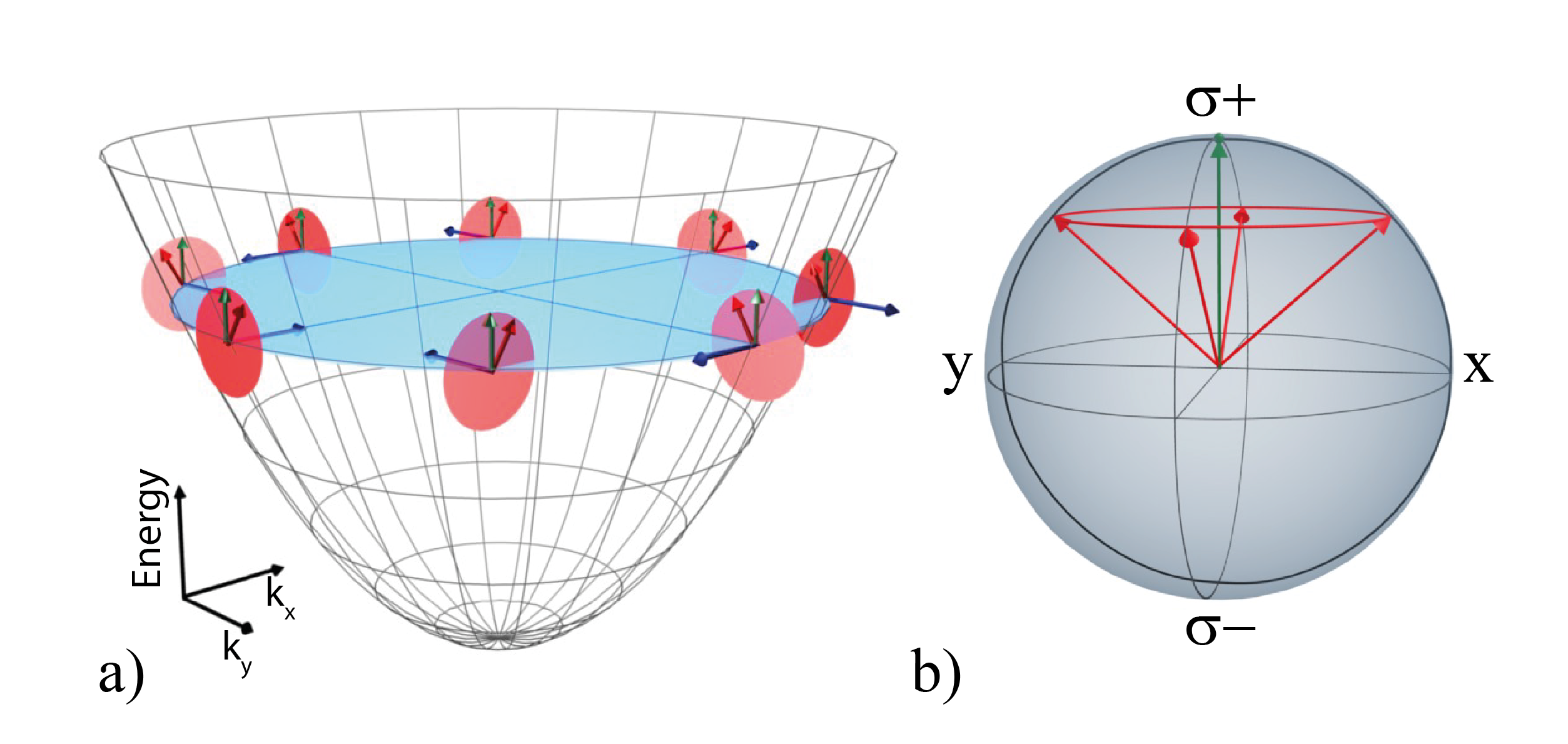}
% Here is how to import EPS art
\caption{(a) Illustration of the k-dependant effect of the effective
magnetic field (TE-TM splitting) on the Stokes vector of the photons (in blue: effective magnetic field; in green: initial Stokes vector orientation; in red: rotated Stokes vector). The parabolic dispersion of the cavity mode is sketched in black solid lines. (b) Mapping of the Stokes vectors to the Poincar\'e sphere (same color code as in (a)).}
\label{fig:figure0}
\end{figure}

In this work, we demonstrate that conversion of spin into orbital angular momentum can also be achieved, in the linear regime, with a planar semiconductor microcavity~\cite{Kavokin2007} -- a compact, layered nanostructure, acting as a single optical element. At first sight, such conversion is unexpected since a semiconductor microcavity is a planar isotropic system. However, microcavities exhibit a polarization splitting between TE and TM polarized modes~\cite{Panzarini1999}, which can be represented by an effective magnetic field whose direction depends on the reciprocal space position excited~\cite{Kavokin2005} despite the isotropic nature of the microcavity (Fig.\ref{fig:figure0}(a)). This directional dependence arises from the choice of a fixed coordinate system when defining of the TE-TM basis, and it does not represent a breaking of the rotational symmetry of the system. This effective magnetic field is responsible for polarized pattern formation~\cite{Langbein2007}, the ``all-optical'' spin Hall effect~\cite{Maragkou2010} and has been predicted to allow conversion between spin and orbital angular momentum~\cite{Liew2007}. The effect of the TE-TM splitting on a circularly polarized distribution can be understood intuitively from Fig.\ref{fig:figure0}, which shows the Stokes vectors for light excited on a ring in reciprocal space (the Stokes vectors represent the polarization state of a light mode on the Poincar\'e sphere as shown in Fig.\ref{fig:figure0}(b)). Whilst a fully circularly polarized distribution would be characterized by Stokes vectors pointing in the vertical direction (green arrows in Fig.\ref{fig:figure0}), their precession about the effective magnetic fields causes them to evolve in different directions. A cross-circularly polarized component develops and the phase profile can become dependant on the angle, as we will show. The process conserves the total angular momentum (spin + orbital angular momentum) via spin-to-orbital angular momentum (SOAM) conversion. We provide a theoretical model that is able to quantitatively reproduce the experimental observations.
\begin{figure}[tb]
\includegraphics[width=0.5\textwidth]{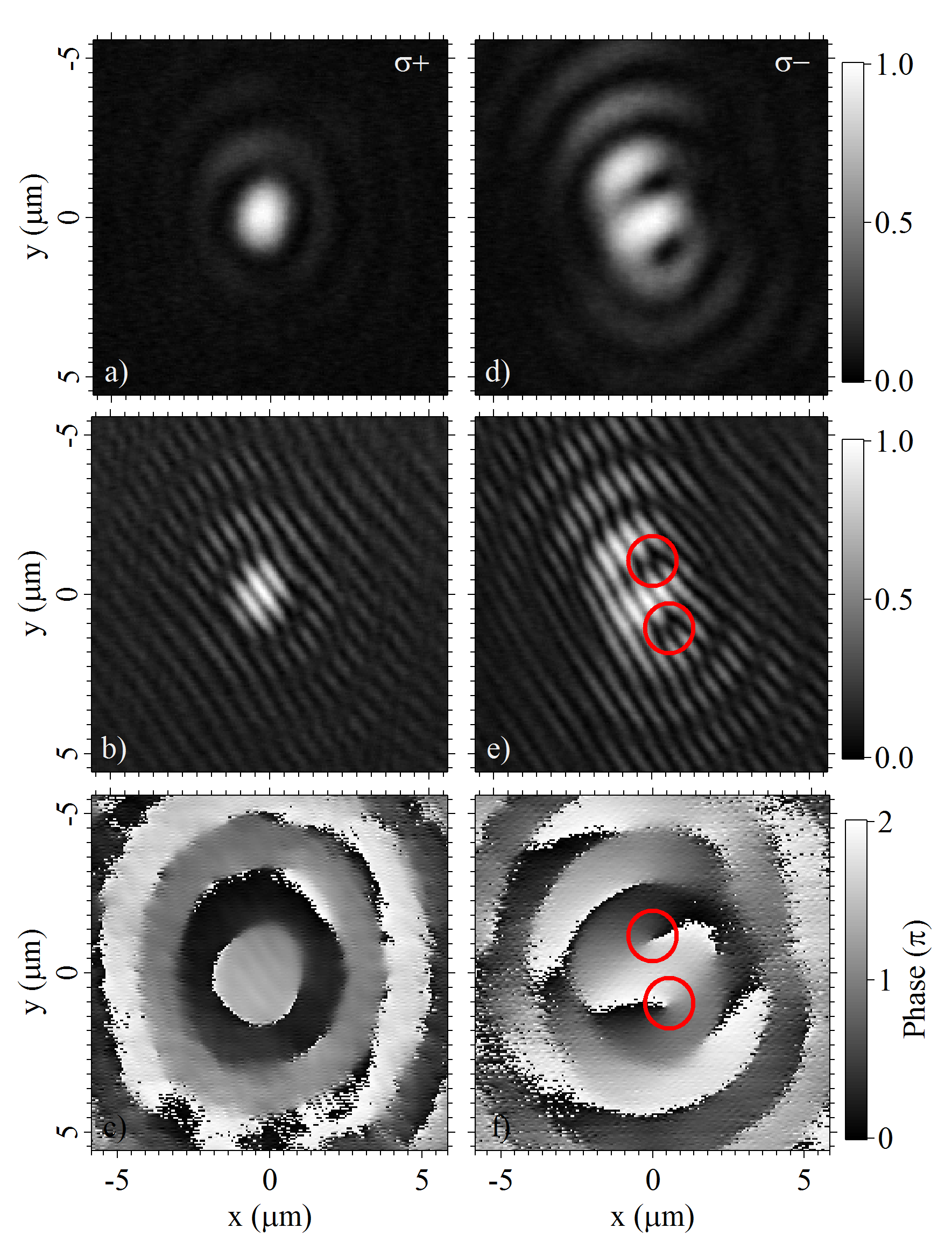}
% Here is how to import EPS art
\caption{Real space profile (a) of the transmitted optical field,
interferogram (b) and corresponding phase (c) for the $\protect\sigma_+$
polarization. Same for the $\protect\sigma_-$ polarization in (d), (e) and
(f) respectively. The red circles mark the position of the pair of optical
half-vortices.}
\label{fig:figure1}
\end{figure}

Formally, under a resonant optical excitation, the dynamics of the intra-cavity field is given by the two component Schr\"{o}dinger equation:
\begin{align}
i\hbar \frac{\partial \psi _{\pm }(\mathbf{k},t)}{\partial t}& =E(k)\psi
_{\pm }(\mathbf{k},t)+\left( \frac{\Delta (k)}{2}e^{\mp 2i\phi }+\frac{\xi }{%
2}\right) \psi _{\mp }(\mathbf{k},t)  \notag \\
& \hspace{10mm}+f_{\pm }(\mathbf{k},t),  \label{eq:Schrodinger}
\end{align}%
where $\pm $ represents the two circularly polarized components of the field $\psi _{\pm}(\mathbf{k},t)$. The 2D in-plane wavevector $\mathbf{k}=(k,\phi)$ is divided into a radial component, $k$, and angular component, $\phi $. $E(k)=E_{c}-i\hbar \Gamma$ represents the complex in-plane dispersion of the cavity mode, where the imaginary component accounts for the decay of photons with rate $2\Gamma $. $\Delta (k)$ and $\xi$ represent the strength of the k-dependent TE-TM polarization splitting~\cite{Panzarini1999} and a possible additional splitting due to anisotropy~\cite{Amo2009}, respectively. The anisotropy term is needed to take into account the small residual linear polarization splitting of real samples. The angular dependence of the TE-TM splitting is equivalent to that represented by the effective magnetic field in Fig \ref{fig:figure0}(a) Ref.~\cite{Kavokin2005}. The pumping term can be written as~\cite{Amo2009}:
\begin{eqnarray}
f_{\pm }(\mathbf{r},t) &=&A_{\pm }e^{-r^{2}/L^{2}}e^{-i\beta
r^{2}}e^{-iE_{p}t/\hbar },  \notag \\
f_{\pm }(\mathbf{k},t) &=&\int f_{\pm }(\mathbf{r},t)\frac{i\Gamma
e^{-iE_{p}t/\hbar }}{E(k)-E_{p}}d\mathbf{r}e^{i\mathbf{k}\cdot \mathbf{r}},
\end{eqnarray}%
which corresponds to a monochromatic focused Gaussian beam with amplitudes $A_{\pm }$, spot size $L$ and energy $E_{p}$ ($e^{-i\beta r^{2}}$ accounts for the wavefront curvature). Considering a circularly polarized pump with $A_{-}=0$, the steady state solution of Eq.~\ref{eq:Schrodinger} is given by the equations:
\begin{eqnarray}
\psi _{-}(\mathbf{k},t) &=&-\frac{\Delta (k)e^{2i\phi }+\xi }{E(k)}\psi _{+}(%
\mathbf{k},t)  \label{eq:psiminus} \\
\psi _{+}(\mathbf{k},t) &=&-\frac{f_{+}(k,t)}{E(k)-\frac{\left( \Delta
(k)e^{-2i\phi }+\xi \right) \left( \Delta (k)e^{2i\phi }+\xi \right) }{E(k)}}
\label{eq:psiplus}
\end{eqnarray}

In the limit $\xi\mapsto0$, where the anisotropy term is neglected, Eq.~\ref{eq:psiplus} becomes independent of $\phi$, that is, the $\sigma_+$ polarized cavity field (the same polarization as the pump) does not carry
orbital angular momentum. However, it is seen from Eq.~\ref{eq:psiminus} that the $\sigma_-$ component carries orbital angular momentum due to the phase profile $e^{2i\phi}$. This phase winding structure corresponds to vortical entities in the optical transmitted signal that are associated with the generation of an orbital angular momentum. The presence of anisotropy, although not responsible for the SOAM conversion process, can give rise to more complex and exotic structures in the phase profile, as it will be shown below.

To observe the SOAM conversion effect, we perform transmission measurements using the cavity mode of an InGaAs/GaAs microcavity, the same sample used in \cite{Paraiso_2009}. Note that although an excitonic resonance exists in the sample, it has negligible effect on the cavity mode resonance, since we operate in the regime of very large positive detuning. We excite in one circular polarization and we detect both circular components of the transmitted signal. In order to demonstrate the existence of vortex excitations we implement a homodyne detection scheme in the setup, which allows to retrieve the phase of the signal~\cite{Nardin2010}. The sample is kept in a liquid helium flow cryostat at low temperature ($\approx 4$ K). The laser source used here is a tunable single-mode CW Ti:Sapphire laser. The laser is first split into two beams: one is used as the phase reference for the homodyne detection, whilst the other is prepared in the circular polarization state $\sigma _{+}$ and is used to excite the system. This gaussian-shaped $\sigma _{+}$ beam is tightly focused onto the sample using a $0.5$ NA microscope objective. The tight spatial focusing results in a very broad, single-energy excitation in momentum space. Consequently, increasing the excitation energy with respect to the bottom of the cavity mode, we are able to excite a ring with $|\mathbf{k}|\neq 0$ in reciprocal space. This is a key feature that allows us to access a regime where the k-dependent TE-TM splitting becomes significant. The real-space transmitted field coming from the sample is collected with another $0.5$ NA objective. The two cross-circular polarizations ($\sigma _{+}$, $\sigma _{-}$) are eventually separated and imaged on a CCD camera using a Wollaston prism. On the same camera we superimpose the homodyne reference beam in order to obtain interference patterns, from which we extract the phase profiles.

For an excitation energy $2.5$ meV above the bottom of the cavity mode, the ring diameter in reciprocal space has a radius of $|\mathbf{k}|\approx 2 \mu m ^{-1}$ for which the TE-TM splitting is estimated to be $\Delta(k)\approx 20\mu eV$. The polarization resolved transmitted optical field is shown in Fig.~\ref{fig:figure1}(a) for $\sigma_+$ and in Fig.~\ref{fig:figure1}(d) for $\sigma_-$ polarizations. The transmitted $\sigma_+$ beam is characterized by a central high-intensity spot surrounded by a series of rings. The cross-polarized $\sigma_-$ signal shows a much wider central spot, featuring two distinct local minima, located at $(x,y)\approx(0.5,-1)\mu m$ and $(x,y)\approx(0.5,1)\mu m$. These minima are attributed to optical vortices because of the SOAM conversion process.

By interfering the transmitted signal with the reference beam, two interferograms are obtained: one for the $\sigma_+$ and one for the $\sigma_-$ transmitted beams, shown in Figs.~\ref{fig:figure1}(b) and (e) respectively. Within the central spot of the $\sigma_-$ signal we unambiguously identify the presence of two fork-like dislocations, highlighted by the red circles in (Fig.~\ref{fig:figure1}(e)). These dislocations coincide with the local density minima of Fig.~\ref{fig:figure1}(d), proving the existence of quantized vortices.
Note that both interferograms (Figs.~\ref{fig:figure1}(b) and (e)) feature other fork-like dislocations, located at the outer edge of the central spot. These dislocations result from the $\pi$ phase mismatch (see Fig.\ref{fig:figure1}(c)) between the central region and the first density ring and, in fact, do not correspond to optical vortices.

Using digital holography methods~\cite{lagous2009}, we extract the phase of the transmitted field for both $\sigma_+$ and $\sigma_-$ polarizations as shown in Figs.~\ref{fig:figure1}(c) and (f) respectively. The analysis of the phase confirms that no topological charge is present in the $\sigma_+$ transmitted signal (Fig.~\ref{fig:figure1}(c)). Nevertheless we observe the aforementioned phase structure associated to the ring density profile of Fig.~\ref{fig:figure1}(a): each ring is in anti-phase with the one before and after. This $\pi$ phase shift is responsible for the fork-like dislocations that can be found at the locations $(x,y)\approx(-1.5,1)\mu m$ and $(x,y)\approx(1,-1)\mu m$ in the interferogram of Fig.~\ref{fig:figure1}(b), which do not correspond to vortices (as they do not coincide with any local density minima).

In the $\sigma_-$ transmitted signal, the vortex singularities exhibit a linear phase increase as a function of the azimuthal angle, thus confirming that we are in the presence of a pair of optical vortices. The total phase jump along the $\sigma_-$ outer transmission ring is found to be $4\pi$, as in the theoretical prediction (see Eq.~\ref{eq:psiminus}). The two vortices present in the $\sigma_-$ transmission are the signature of the $L=+2$ orbital angular momentum obtained by the conversion of the spin, from $\sigma_+$ ($S=+1$) to $\sigma_-$ ($S=-1$) due to the TE-TM polarization splitting.

Let us now comment on the topological charge of the observed vortices: up to now, we have considered these singularities as singly quantized vortices present in the $\sigma _{-}$ transmitted signal and we have treated the $\sigma _{+}$ transmission as an independent spin. It is of course worth noting that there is complete mutual coherence between the $\sigma _{+}$ and the $\sigma _{-}$ polarizations as the latter is the product of a fully
coherent process. It is therefore necessary to account for the flat phase and non vanishing density of the $\sigma _{+}$ polarization component when considering the $2\pi$ phase shift and density minimum in $\sigma _{-}$, which forcedly renders the observed vortical entities a clear signature of unconventional vorticity. Indeed, the topological entities that we observe correspond to optical half quantum vortices (HQVs)~\cite{Rubo2007, lagous2009}, although here their position can be controlled directly by the laser position since we use a resonant excitation.

Theory predicts that the two optical half-vortices resulting from SOAM conversion, are expected to spatially overlap in the absence of anisotropy. When anisotropy is present, the pair of optical HQVs is spatially separated by a distance directly related to the degree of anisotropy in the sample. This anisotropy being strongly
position dependent, allowed us to find positions where the two half-vortices were closer or farther apart from each other. In general, though, the SOAM conversion is robust and generic on the sample surface.

Finally, to further verify that the observed singularities are a product of the SOAM conversion, we performed the same experiment exciting the parabolic dispersion resonant to $k = 0$ where the TE-TM splitting is $\Delta(0)=0\mu eV$. We did not encounter any phase singularities neither in the $\sigma_+$ nor the $\sigma_-$ transmitted signals thus verifying the dependence of the SOAM conversion on the TE-TM polarization splitting.

\begin{figure}[tb]
\includegraphics[width=0.5\textwidth]{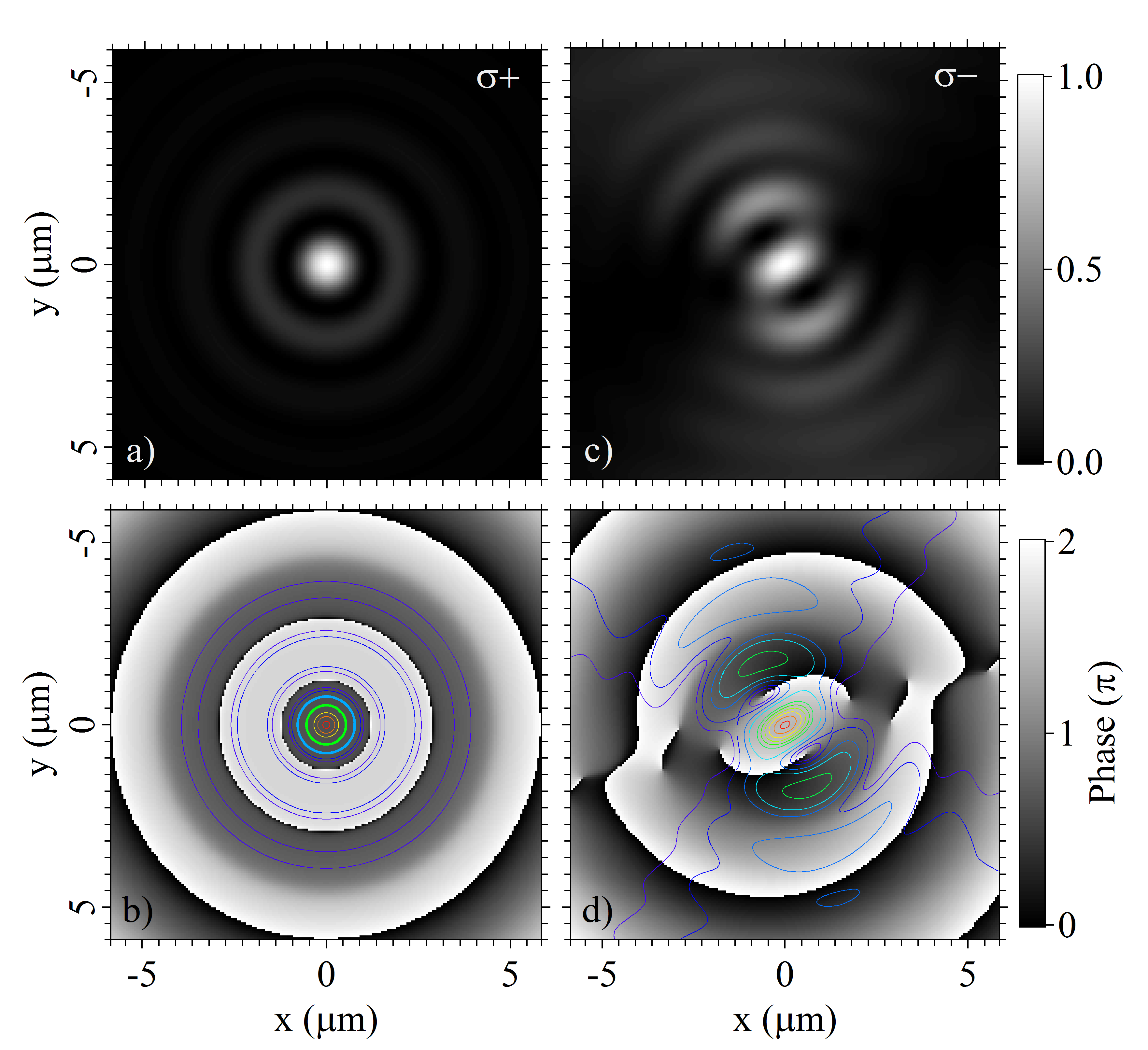}
% Here is how to import EPS art
\caption{Real space intensity profile of the intra-cavity optical field and
corresponding phase: (a,b) for the $\protect\sigma_+$ and (c,d) $\protect%
\sigma_-$ polarizations, respectively. Note: the density contour lines of (a,c) are superposed to the corresponding phase in (b,d) in order to underline that only two local density minima are present, in the $\sigma_-$ signal, and that they match with two phase windings of $2\pi$. The presence of anisotropy is responsible for the complex and exotic phase structure.}
\label{fig:figure2}
\end{figure}

We are able to quantitatively reproduce the experimental findings with the theoretical model introduced above~\cite{simulpar}. The results of the simulations are shown in Fig.~\ref{fig:figure2}. The
intensity of the optical field is plotted together with the corresponding calculated phase, for the $\sigma_+$ in Fig.~\ref{fig:figure2}(a,b) and for the $\sigma_-$ polarization in Fig.~\ref{fig:figure2}(c,d), respectively. As it can be seen in those figures, the theory captures all the features of the experimental findings. In the model, the spatial separation between the two half-vortices is caused by the $\xi$ anisotropy term.

In this work we investigated the creation of orbital angular momentum carrying beams exploiting an optical microcavity, a 2D optical isotropic and homogeneous element. Our system represents a single compact optical device capable of generation of beams with orbital angular momentum for applications in the control of nanomachines or quantum information. Moreover, being a homogeneous and isotropic optical system, the creation of arrays of OAM carrying elements is greatly simplified, not requiring any special nanostructuring. We have demonstrated the creation of a pair of optical half-vortices, forcing the $\sigma _{-}$ beam to carry a total angular momentum of $+2$. This observation proves a complete SOAM conversion occurring between the two orthogonal circular polarizations, which are coupled by the TE-TM splitting featured by the microcavity sample. With the theoretical model presented, we are able to fit the experimental parameters and quantitatively match the results.

This work was supported by the Swiss National Science Foundation through NCCR ``Quantum Photonics''.


\begin{thebibliography}{99}

\bibitem{FrankeArnold2008} S. Franke-Arnold, L. Allen, M. Padgett, Laser \& Photon. Rev., \textbf{2}, 299 (2008).

\bibitem{Beth1936} R. A. Beth, Phys. Rev., \textbf{50}, 115 (1936).

\bibitem{He1995} H. He, M. E. J. Friese, N. R. Heckenberg, H. Rubinsztein-Dunlop, Phys. Rev. Lett., \textbf{75}, 826 (1995).

\bibitem{Grier2003} D. G. Grier, Nature, \textbf{424}, 810 (2003).

\bibitem{Garetz1997} B. A. Garetz \& S. Arnold, Opt. Commun., \textbf{31}, 1 (1997).

\bibitem{Courtial1998} J. Courtial, K. Dholakia, L. Allen, M. J. Padgett, Phys. Rev. A, \textbf{56}, 4193 (1997).

\bibitem{Andersen2006} M. F. Andersen, C. Ryu, P. Clad\'{e}, V. Natarajan, A. Vaziri, K. Helmerson, W. D. Phillips, Phys. Rev. Lett., \textbf{97}, 170406 (2006).

\bibitem{Mair2001} A. Mair, A. Vaziri, G. Weihs, A. Zeilinger, Nature, \textbf{412}, 313 (2001).

%\bibitem{FrankeArnold2002} S. Franke-Arnold, Barnett S. M., M. J. Padgett, L. Allen, Phys. Rev. Lett., \textbf{65}, 033823 (2002).

\bibitem{Granata2010} M. Granata, C. Buy, R. Ward, M. Barsuglia, \textbf{105},
231102 (2010).

\bibitem{Allen1992} L. Allen, M. W. Beijersbergen, R. J. C. Spreeuw, J. P. Woerdman, Phys. Rev. A, \textbf{45}, 8185 (1992).

\bibitem{Oemrawsingh2004} S. S. R. Oemrawsingh, E. R. Eliel, J. P. Woerdman, E. J. K. Verstegen, J. G. Kloosterboer, G. W. t'Hooft, J. Opt. A: Pure Apply. Opt., \textbf{6}, S288 (2004).

\bibitem{Beijersbergen1994} M. W. Beijersbergen, R. P. C. Coerwinkel, M. Kristensen, J. P. Woerdman, Opt. Commun., \textbf{112}, 321 (1994).

\bibitem{Bazhenov1990} V. Yu. Bazhenov, M. V. Vasnetsov, M. S. Soskin, JETP Lett., \textbf{52}, 429 (1990).

\bibitem{Marrucci2006} L. Marrucci, C. Manzo, D. Paparo, Phys. Rev. Lett., \textbf{96}, 163905 (2006).

\bibitem{Brasselet2009} E. Brasselet, N. Murazawa, H. Misawa, S. Juodkazis, Phys. Rev. Lett., \textbf{103}, 103903 (2009).

\bibitem{Zhao2007} Y. Zhao, J. S. Edgar, G. D. M. Jeffries, D. McGloin, D. T. Chiu, Phys. Rev. Lett., \textbf{99}, 073901 (2007).

\bibitem{Kavokin2007} A. V. Kavokin, J. J. Baumberg, G. Malpuech, F. P. Laussy, \textit{Microcavities}, Oxford University Press (2007).

\bibitem{Panzarini1999} G. Panzarini, L. C. Andreani, A. Armitage, D. Baxter, M. S. Skolnick, V. N. Astratov, J. S. Roberts, A. V. Kavokin, M. R. Vladimirova, M. A. Kaliteevski, Phys. Rev. B, \textbf{59}, 5082 (1999).

\bibitem{Kavokin2005} A. V. Kavokin, G. Malpuech, M. M. Glazov, Phys. Rev. Lett., \textbf{95}, 136601 (2005).

\bibitem{Langbein2007} W. Langbein, I. Shelykh, D. Solnyshkov, G. Malpuech, Y. Rubo, A. Kavokin, Phys. Rev. B, \textbf{75}, 075323 (2007).

\bibitem{Maragkou2010} M. Maragkou, C. Richards, T. Ostatnicky, A. J. D. Grundy, J. Zajak, W. Langbein, P. Lagoudakis, \textit{to be published}.

\bibitem{Liew2007} T. C. H. Liew, A. V. Kavokin, I. A. Shelykh, Phys. Rev. B, \textbf{75}, 241301 (2007).

\bibitem{Amo2009} A. Amo, T. C. H. Liew, C. Adrados, E. Giacobino, A. V. Kavokin, A. Bramati, Phys. Rev. B, \textbf{80}, 165325 (2009).

\bibitem{Paraiso_2009} T. K. Para\"iso, D. Sarchi, G. Nardin, R. Cerna, Y. Leger, B. Pietka, M. Richard, O. El Da\"if, F. {Morier-Genoud}, V. Savona, B. Deveaud-Pl\'edran, Phys. Rev. B, \textbf{79}, 045319 (2009).

\bibitem{Nardin2010} G. Nardin, Y. Leger, B. Pietka, F. Morier-Genoud, B. Deveaud-Pl\'edran, Phys. Rev. B, \textbf{82}, 045304 (2010).

\bibitem{lagous2009} K. G. Lagoudakis, T. Ostatnicky, A. V. Kavokin, Y. G. Rubo, R. Andr\'e, B. Deveaud-Pl\'edran, Science, \textbf{326}, 974 (2009).

\bibitem{Rubo2007} Yu. G. Rubo, Phys. Rev. Lett., \textbf{99}, 106401 (2007).

\bibitem{simulpar} Parameters for the simulations: $E_{up}(k)$ was taken as a parabolic dispersion with effective mass $5\times10^{-5}$ of the free electron mass. The pump energy was chosen as $E_p=E_{up}(0)+3$meV, which excites a ring in k-space at approximately $2\mu m^{-1}$ radius. $\Delta(k)$ was taken as parabolic with $\Delta(2\mu m^{-1})=0.02$meV. The other parameters were: $\xi=0.02$meV, $L=5.1\mu m$, $\Gamma=0.33$meV, $\beta=0.14\mu m^{-2}$.
    
\end{thebibliography}
\end{document}